\documentstyle[fleqn,espcrc2,epsf]{article}
\title{
Non-perturbative renormalization factors of bilinear quark operators
for Kogut-Susskind fermions and light quark masses in quenched QCD
\thanks{presented by N. Ishizuka}
}
\author{
JLQCD Collaboration : 
S.~Aoki
\address{
Institute of Physics, University of Tsukuba,
Tsukuba, Ibaraki 305-8571, Japan
},
M.~Fukugita
\address{
Institute for Cosmic Ray Research,
University of Tokyo,
Tanashi, Tokyo 188-8502, Japan
},
S.~Hashimoto
\address{
High Energy Accelerator Research Organization (KEK),
Tsukuba, Ibaraki 305-0801, Japan
},
K-I.~Ishikawa
\address{
Department of Physics, Hiroshima University,
Higashi-Hiroshima, Hiroshima 739-8526, Japan
},
N.~Ishizuka$^{\rm a,}$
\address{
Center for Computational Physics,
University of Tsukuba,
Tsukuba, Ibaraki 305-8577, Japan
},
Y.~Iwasaki$^{\rm ~a,e}$
K.~Kanaya$^{\rm ~a,e}$,
T.~Kaneda$^{\rm ~a}$,
S.~Kaya$^{\rm ~c}$,
Y.~Kuramashi$^{\rm ~c}$,
M.~Okawa$^{\rm ~c}$,
T.~Onogi$^{\rm ~d}$,
S.~Tominaga$^{\rm ~c}$,
N.~Tsutsui$^{\rm ~d}$,
A.~Ukawa$^{\rm ~a,e}$,
N.~Yamada$^{\rm ~d}$,
T.~Yoshi\'{e}$^{\rm ~a,e}$
}
\begin{document}
%
%
\begin{abstract}
%
Light quark masses are computed for Kogut-Susskind fermions by
evaluating non-perturbatively the renormalization factor for bilinear 
quark operators.
Calculations are carried out in the quenched approximation
at $\beta=6.0$, $6.2$, and $6.4$.
For the average up and down quark mass we find
$m_{\overline{\rm MS}}(2 {\rm GeV})= 4.15(27) {\rm MeV}$
in the continuum limit, which is significantly larger than
$3.51(20) {\rm MeV}$ ($q^*=1/a$)
or $3.40(21) {\rm MeV}$ ($q^*=\pi/a$)
obtained with the one-loop perturbative renormalization factor.
\end{abstract}
\maketitle
%
%
\section{ Introduction }
Light quark masses are important unknown parameters of the standard model,
and a number of lattice QCD calculations have been carried out to evaluate quark masses
employing the Wilson, clover or Kogut-Susskind (KS)
fermion action~\cite{review}.
Among the results, those with the KS action appear more
accurate than others because of small lattice discretization errors
and small statistical errors.

A worry with the KS result, however, has been that
the employed one-loop renormalization factor 
takes a large value of $\approx 2$ in the range of $\beta$ studied,
calling into question the viability of perturbation theory.
In this article we report a study to circumvent this problem:
we calculate the renormalization factor of bilinear quark operators for the
KS action non-perturbatively using the method of Ref.~\cite{martinelli}
developed for the Wilson/clover actions.  This calculation is carried out
in quenched QCD at $\beta=6.0$, $6.2$, and $6.4$ on an $32^4$ lattice.
The results, combined with our previous calculation of bare quark
masses~\cite{JLQCD_M},
lead to a non-perturbative determination of the light quark mass.
%
%
\section{ Method }
The renormalization factor of a bilinear operator ${\cal O}$
is obtained from the amputated Green function,
\begin{equation}
  \Gamma_{\cal O}(p)
= S(p)^{-1} \langle 0 | \phi ( p ) {\cal O} \bar{\phi} (p) | 0 \rangle
  S(p)^{-1}
\end{equation}
where the quark two-point function is defined by
$S(p)=\langle 0 | \phi ( p ) \bar{\phi} (p) | 0 \rangle$.
The quark field $\phi (p)$ with momentum $p$ is defined from the
original one-component field $\chi(x)$ by
$\phi_A (p) = \sum_y {\rm exp}( - i p\cdot y) \chi ( y + aA )$,
where $y_\mu=2a n_\mu$, $p_\mu = 2\pi/(aL) n_\mu$ ($n_\mu=[-L/4,L/4)$) and
$A_\mu=[0,1]$.

The renormalization condition imposed on $\Gamma_{\cal O}(p)$ is given by
\begin{equation}
   \Gamma_{\cal O}(p) =   Z_{\phi}(p) Z_{\cal O}(p) \Gamma_{\cal O}^{(0)}
\label{eq:RI}
\end{equation}
where $\Gamma_{\cal O}^{(0)}$ is the amputated Green function at tree level
and $Z_{\phi}(p)$ is the wave function renormalization factor
which can be calculated by the condition $Z_V(p)=1$
for the conserved vector current.

The relation between the bare operator on the lattice
and the renormalized operator in the continuum takes the form,
\begin{equation}
     {\cal O}_{\overline{\rm MS}}(\mu)
= U_{\overline{\rm MS}}(\mu,p) Z_{\overline{\rm MS}}(p) / Z_{\cal O}(p)
    {\cal O}^{\rm lat.}(a)
\label{ZZZ}
\end{equation}
where $U_{\overline{\rm MS}}(\mu,p)$ is the renormalization-group running factor,
and $Z_{\overline{\rm MS}}(p)$ is the matching factor from the
${\rm RI}$ scheme defined by (\ref{eq:RI})
to the $\overline{\rm MS}$ scheme, calculated perturbatively in the continuum.
For the light quark mass we apply relation (\ref{ZZZ}) in the
scalar channel in the chiral limit.

We use a source in momentum eigenstate to evaluate quark propagators.
This results in very small statistical errors of $O(0.1\%)$
in the Green functions.

The external momentum $p$ should be taken in the range
$\Lambda_{\rm QCD} \ll p \ll O(1/a)$
in order to keep under control higher order effects
in continuum perturbation theory,
non-perturbative hadronization effect on the lattice,
and discretization errors on the lattice.
In this work we choose $15$ momenta in the range $0.038553 < (ap)^2 < 1.9277$
for all values of $\beta$.
%
%
\begin{figure}[t]
\vspace*{-0.7cm}
\centerline{\epsfxsize=7.0cm \epsfbox{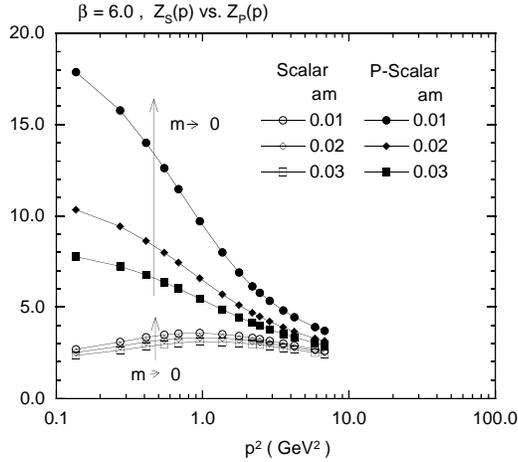}}
\vspace*{-1.0cm}
\caption{\label{FIG1}
The scalar renormalization factor $Z_S(p)$ 
and that for the pseudoscalar $Z_P(p)$.
}
\vspace*{-0.7cm}
\end{figure}
%
%
\begin{figure}[t]
\vspace*{-0.7cm}
\centerline{\epsfxsize=7.0cm \epsfbox{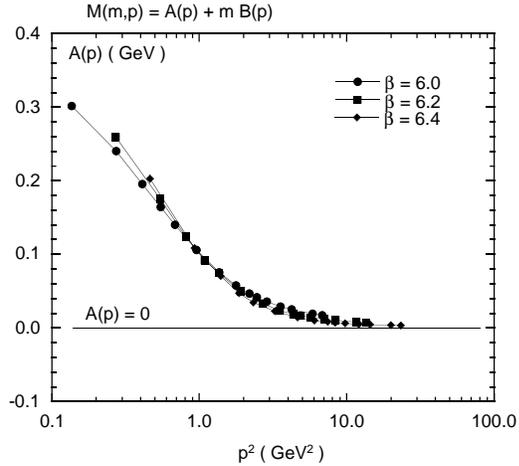}}
\vspace*{-1.0cm}
\caption{\label{FIG5}
$M(p)$ in the chiral limit.
}
\vspace*{-0.7cm}
\end{figure}
%
%
\section{ Result }
In Fig.~\ref{FIG1} we compare the scalar renormalization factor $Z_S (p)$
with that for pseudoscalar $Z_P(p)$ for three values of bare quark mass $a m$
at $\beta=6.0$.
From chiral symmetry of KS fermions,
we expect naively $Z_S(p)= Z_P(p)$ for all momenta $p$ in the chiral limit.
Clearly this relation does not hold with our result toward small momenta,
where $Z_P(p)$ rapidly increases as $m\to 0$,
while $Z_S(p)$ does not show such a trend.

To understand this result, we note that chiral symmetry of KS fermion
leads to the following identities :
\begin{eqnarray}
 &&  Z_S(p) \cdot Z_\phi(p) = { \partial M(p) } / { \partial m }   \cr
 &&  Z_P(p) \cdot Z_\phi(p) = {  M(p) } / {  m }
\label{ZP}
\end{eqnarray}
with $M(p) = {\rm Tr}[ S(p)^{-1} ]$.
In Fig.~\ref{FIG5} $M(p)$ in the chiral limit obtained
by a linear extrapolation in $m$ is plotted.
It rapidly dumps for large momenta,
but takes large values in the small momentum region.
Combined with (\ref{ZP}) this implies that $Z_P(p)$
diverges in the chiral limit for small momenta, which is consistent with
the result in Fig.~\ref{FIG1}.

The function $M(p)$ is related to chiral condensate as follows :
\begin{equation}
   \langle \phi \bar{\phi} \rangle = \sum_p {\rm Tr}[ S(p) ]
    = \sum_p \frac{ M(p) }{ C_\mu(p)^2 + M(p)^2 }
\end{equation}
where $C_\mu(p) = -i {\rm Tr}[ (\gamma_\mu \otimes I ) S(p)^{-1} ]/ \cos(p_\mu a)$.
A non-vanishing value of $M(p)$ for small momenta would lead to
a non-zero value of the condensate.
Therefore the divergence of $Z_P (p)$ near the chiral limit
is a manifestation of spontaneous symmetry breakdown of chiral symmetry;
it is a non-perturbative effect
arising from the presence of massless Goldstone boson.

We expect this non-perturbative effect to affect
the scalar renormalization factor $Z_S(p)$ much less,
since the scalar operator can not interact
directly with the pseudoscalar meson.
Indeed the quark mass dependence is quite small as we have seen 
in Fig.~\ref{FIG1}.
%
%
\begin{figure}[t]
\vspace*{-0.7cm}
\centerline{\epsfxsize=7.0cm \epsfbox{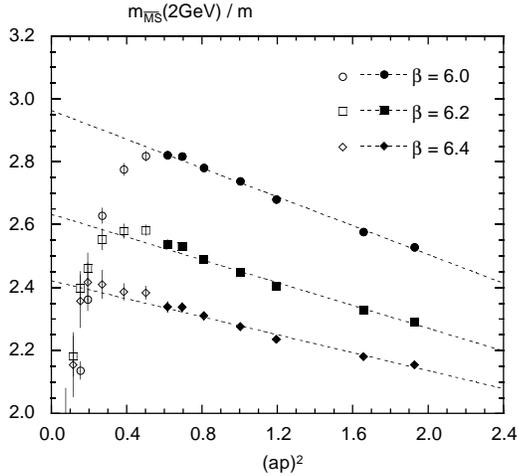}}
\vspace*{-1.0cm}
\caption{\label{FIG8}
The ratio $m_{\overline{\rm MS}}(\mu) / m$
at $\mu=2{\rm GeV}$.
For each $\beta$ the filled data points are used
for linear extrapolation in $(ap)^2$.
}
\vspace*{-0.7cm}
\end{figure}
%
%

In Fig.~\ref{FIG8} we show the momentum dependence of the ratio
$m_{\overline{\rm MS}}(\mu) / m
= U_{\overline{\rm MS}}(\mu,p) Z_{\overline{\rm MS}}(p) Z_S(p)$
calculated in the chiral limit
where we set $\mu=2{\rm GeV}$ and use the three-loop formula~\cite{three_loop}
for $U_{\overline{\rm MS}}$ and $Z_{\overline{\rm MS}}$.
While the ratio should be independent of the quark momentum $p$,
our results show a large momentum dependence which is
almost linear in $(ap)^2$ for $0.6 < (ap)^2$.

A natural origin of the linear dependence on $(ap)^2$
is the lattice discretization error of the scalar operator,
which differs by terms of $O(a^2)$ from that of the continuum for the KS fermion.
We then remove this error from the renormalization factor by
a linear extrapolation in $(ap)^2$ to $(ap)^2=0$.
In Fig.~\ref{FIG8} the fitting lines are plotted,
where filled data points are used for the linear extrapolation.
For comparison, the ratio calculated with the one-loop value
equals $1.867$, $1.877$, and $1.871$ for $\beta=6.0$, $6.2$ and $6.4$ at $q^*=1/a$.
Hence one-loop perturbation theory underestimates the ratio by $40\%$ to $20\%$.

\begin{figure}[t]
\vspace*{-0.7cm}
\centerline{\epsfxsize=7.0cm \epsfbox{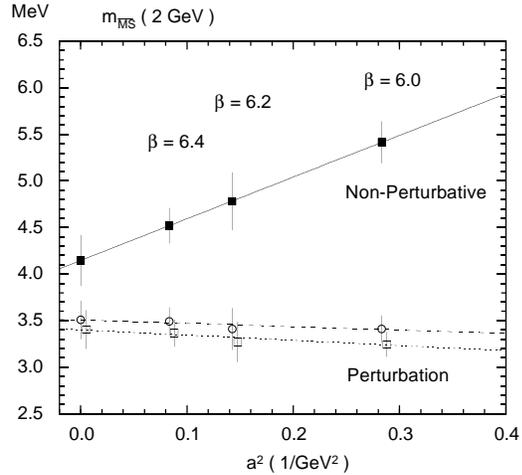}}
\vspace*{-1.0cm}
\caption{\label{FIG9}
The final results of the light quark mass at $\mu=2 {\rm GeV}$.
}
\vspace*{-0.7cm}
\end{figure}

Our final results for the averaged up and down quark mass
at $\mu=2{\rm GeV}$ are shown in Fig.~\ref{FIG9} by filled symbols.
Here we use the JLQCD results for bare quark mass~\cite{JLQCD_M}.
The values are substantially larger than those obtained with one-loop
perturbation theory (open circles for $q^*=1/a$ and squares for $q^*=\pi/a$).
Furthermore they exhibit a significant $a^2$ dependence,
which we ascribe to the discretization error of the quark mass itself.
Making a linear extrapolation in $a^2$,
our final result in the continuum limit is given by
\begin{equation}
  m_{\overline{\rm MS}}(2 {\rm GeV}) = 4.15(27) {\rm MeV}.
\end{equation}
This value is $20\%$ larger than the perturbative estimates :
$3.51(20){\rm MeV}$ for $q^*=1/a$ and $3.40(21){\rm MeV}$ for $q^*=\pi/a$.
%
%

This work is supported by the Supercomputer Project No.32 (FY1998)
of High Energy Accelerator Research Organization (KEK),
and also in part by the Grants-in-Aid of the Ministry of 
Education (Nos.~08640404, 09304029, 10640246, 10640248, 10740107, 10740125).
S.K. and S.T. are supported by the JSPS Research Fellowship.
%
%
%
%
%

%

\vspace*{-0.4cm}
\begin{thebibliography}{9}
%
%
\bibitem{review}
R. Gupta and T. Bhattacharya,
Nucl. Phys. {\bf B}(Proc. Suppl.){\bf 63} (1998) 95.
%
\bibitem{martinelli}
G. Martinelli {\it et.al. }, Nucl. Phys. {\bf B445} (1995) 81.
%
\bibitem{three_loop}
E. Franco and V. Lubicz, hep-ph/9803491;
T. van Ritbergen {\it et.al. }, Phys. Lett. {\bf B400} (1997) 379;
J.A,M, Vermaseren {\it et.al. }, Phys. Lett. {\bf B405} (1997) 327;
K.G. Chetyrkin, Phys. Lett. {\bf B404} (1997) 161;
%
\bibitem{JLQCD_M}
JLQCD Collaboration, Nucl. Phys. {\bf B}(Proc. Suppl.) {\bf 53} (1997) 209.
%
\end{thebibliography}
\end{document}